\documentclass[reprint,prx,notitlepage,superscriptaddress]{revtex4-1}

\usepackage{graphics,epsfig,amsfonts,amssymb,amsmath,color}
\usepackage[normalem]{ulem}
\usepackage{bm,bbm}
\usepackage[mathcal]{euscript}
\usepackage{soul}

\usepackage[monochrome]{xcolor}
\usepackage{float}

\newcommand{\blue}[1]{\textcolor{blue}{#1}}

\colorlet{BLUE}{blue}
\protected\def\myblue{\color{blue}}

\begin{document}

\title{
\blue{Heuristic machinery for thermodynamic studies of SU(N) fermions \\ with neural networks}}
\author{Entong Zhao}
\affiliation{Department of Physics, The Hong Kong University of Science and Technology,\\ Clear Water Bay, Kowloon, Hong Kong, China}
\author{Jeongwon Lee}
\affiliation{HKUST Jockey Club Institute of Advanced Study, The Hong Kong University of Science and Technology,\\ Clear Water Bay, Kowloon, Hong Kong, China}
\author{Chengdong He}
\affiliation{Department of Physics, The Hong Kong University of Science and Technology,\\ Clear Water Bay, Kowloon, Hong Kong, China}
\author{Zejian Ren}
\affiliation{Department of Physics, The Hong Kong University of Science and Technology,\\ Clear Water Bay, Kowloon, Hong Kong, China}
\author{Elnur Hajiyev}
\affiliation{Department of Physics, The Hong Kong University of Science and Technology,\\ Clear Water Bay, Kowloon, Hong Kong, China}
\author{Junwei Liu}
\affiliation{Department of Physics, The Hong Kong University of Science and Technology,\\ Clear Water Bay, Kowloon, Hong Kong, China}
\author{Gyu-Boong Jo}
\affiliation{Department of Physics, The Hong Kong University of Science and Technology,\\ Clear Water Bay, Kowloon, Hong Kong, China}

\begin{abstract}
	The power of machine learning (ML) provides the possibility of analyzing experimental measurements with an unprecedented sensitivity. However, it still remains challenging to probe the subtle effects directly related to physical observables and to understand physics behind from ordinary experimental data using ML. Here, we introduce a heuristic machinery by using machine learning analysis. We use our machinery to guide the thermodynamic studies in the density profile of ultracold fermions interacting within SU($N$) spin symmetry prepared in a quantum simulator. Although such spin symmetry should manifest itself in a many-body wavefuction, it is elusive how the momentum distribution of fermions, the most ordinary measurement, reveals the effect of spin symmetry. Using a fully trained convolutional neural network (NN) with a remarkably high accuracy of $\sim$94~$\%$ for detection of the spin multiplicity, we investigate how the accuracy depends on various less-pronounced effects with filtered experimental images. Guided by our machinery, we directly measure a thermodynamic compressibility from density fluctuations within the single image. Our machine learning framework shows a potential to validate theoretical descriptions of SU($N$) Fermi liquids, and to identify less-pronounced effects even for highly complex quantum matter with minimal prior understanding.\end{abstract}

%
%
%

\maketitle

Multi-component fermions with SU($N$)-symmetric interactions hold a singular position as a prototype system for understanding quantum many-body phenomena in condensed matter physics,  high energy physics and atomic physics ~\cite{Cazalilla:2014kq}. In condensed matter, for example, interacting electrons usually possess SU(2) symmetry, while there are emergent higher spin symmetry for the low-energy property of systems, as the SU($4$) symmetry in graphene due to the combination of spin and valley degrees of freedom~\cite{CastroNeto:2009cl}. In quantum chromodynamics, nuclear interactions are represented by SU($3$) symmetry~\cite{GellMann:1962hl,Neeman:1961ea}. In the past decades, developments in cooling and trapping of alkaline-earth-like (AEL) fermions~\cite{He:2019bb} have opened new possibilities to achieve even higher spin symmetries owing to their distinctive inter-particle interactions, and thus provided ideal platforms to study various SU($N$) fermionic systems~\cite{Cazalilla:2014kq,Cazalilla:2009wl,Zhang:2020jh}. 
Although the role of SU($N$) symmetry has been probed in optical lattices~\cite{Gorshkov:2010hw,Taie:2012tb,Zhang:2014el,Cappellini:2014vo,Scazza:2014jw,Hofrichter:2016iq,Ozawa:2018ge,Goban:2018cq}, the comprehensive characterization of interacting SU($N$) fermions in bulk, wherein the SU($N$) Fermi liquid description is valid, has still remained challenging~\cite{Pagano:2014hy,Song:2019wb,He:2020vm,Sonderhouse:2020wz}. \blue{One of the bottlenecks is that the interaction-induced effect enhanced by enlarged SU($N$) symmetry is sufficiently pronounced by the tight confinement only in 1D~\cite{Pagano:2014hy} or 2D cases~\cite{He:2020vm}. It is only recently that thermodynamics and contact interactions are investigated in 3D SU($N$) fermions~\cite{Song:2019wb,Sonderhouse:2020wz}, but a comprehensive experimental study of SU($N$) fermions still remains to be done.} Developing new experimental techniques or designing new approaches to uncover the \blue{subtle} connection of various \blue{spin multiplicity dependent} properties with the available experimental measurements in SU($N$) interacting fermions is crucial to advance our understanding of SU($N$) symmetry and the corresponding many-body phenomena.

\begin{figure*}[htbp]
	\includegraphics[width=0.9\linewidth]{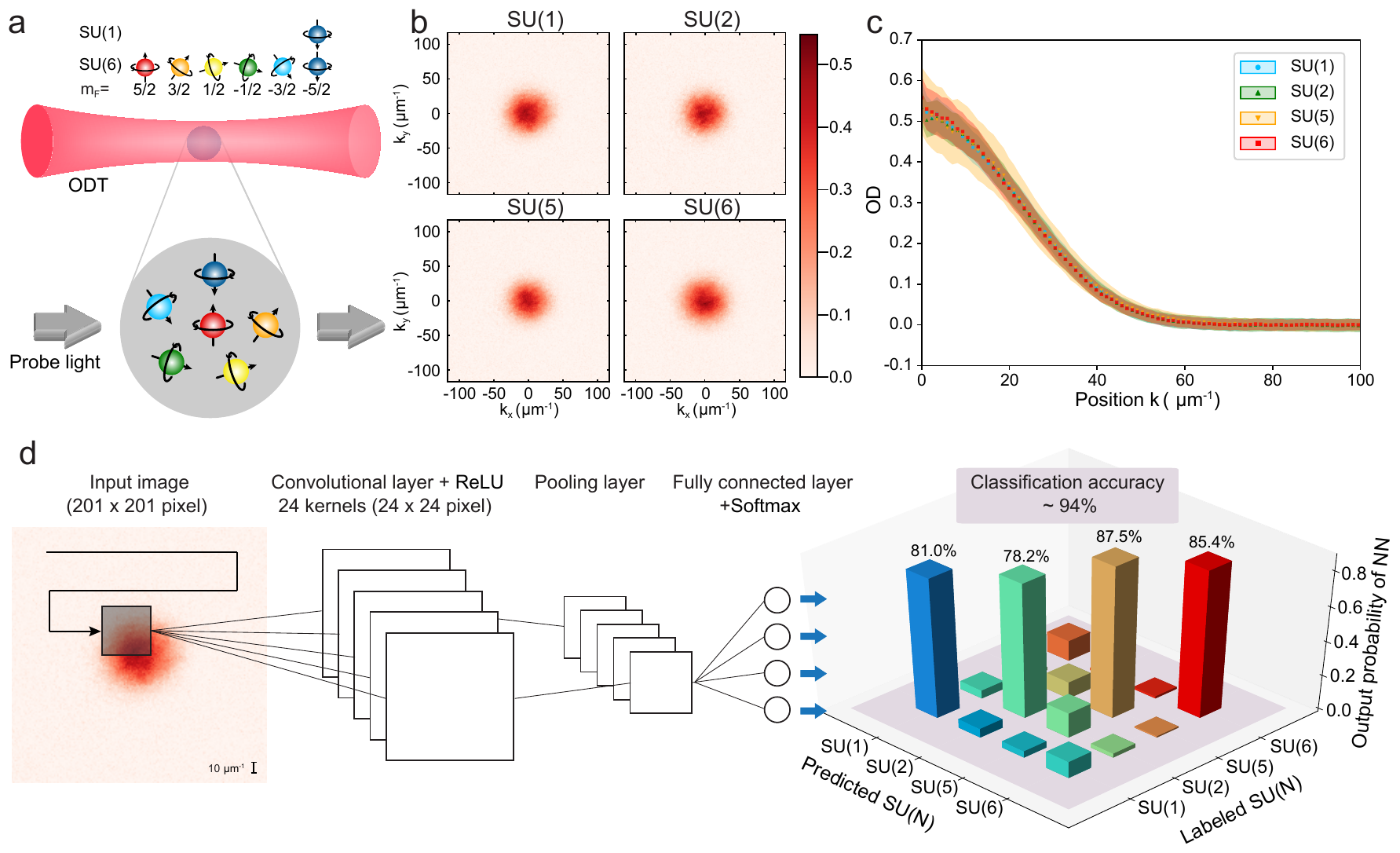}
	\caption{\textbf{Distinguishing SU($N$) fermions based on spin multiplicity by machine learning.} (a) Schematic of preparing SU($N$) gases in the optical dipole trap (ODT).  The momentum distribution of the SU($N$) Fermi liquid of $^{173}$Yb atoms with tunable spin configuration is recorded. The collected datasets are then fed into the NN as the input images for classification. (b) Examples of single experimental images of SU($N$) gases. (c) Radially averaged optical density (OD) profiles in different SU($N$) gases. The shaded region represents the fluctuation of the density profile. (d) Experimental images of SU($N$) gases are loaded into the neural network with one single convolutional layer. The \blue{black} line and window represent how the kernel slides across the image. The output layer classifies the image into one of the class (i.e. SU($1$), SU($2$), SU($5$), SU($6$)) resulting in a classification accuracy around 94\%. For each input image, NN outputs probabilities of different SU($N$) with the highest value of the correct class. The output probabilities of NN are averaged over the test dataset.}\label{fig:1}
\end{figure*}

\blue{Here, we propose a new framework to use machine learning as a guidance for the image analysis in quantum gas experiments and demonstrate the thermodynamic study of SU($N$) fermions. The main idea of this heuristic approach can be summarized into a three-step process: (1) Manually control the amount of information within each of the images we feed to the NNs during the training or testing processes; (2) Determine the relative importance of the given (or removed) information based on the changes in the accuracy of the training or testing processes; (3) Identify the connection between the information and specific physical observables, which we can further focus our analytical efforts on.}


\blue{To demonstrate the proposed machinery concretely, we take a density profile of SU($N$) Fermi gases as an example, and show how it can guide the analytical studies. Besides the pronounced effects such as atom number and fugacity, we explore the connection between the spin multiplicity and less-pronounced features, such as compressibility and Tan's contact in the density profile.}
\blue{Based on this machinery, we demonstrate that one can extract less-pronounced effects even in the most ordinary density profiles, and successfully reveal  thermodynamic features, which depend on the spin multiplicity, from density fluctuations and high-momentum distributions. This allows one to detect the spin multiplicity with a high accuracy $\sim94\%$ in a single snapshot classification of SU($N$) density profiles.}
To further verify the validity of the connection between the \blue{less-pronounced effects} and physical observable, we further measure the thermodynamic compressibility from density fluctuations within a single image benchmarking ML processes, which turns out to be in very good agreement with SU($N$) Fermi liquid descriptions. 
Besides providing general-purpose methods to \blue{extract various less-pronounced effects} and consolidate our understanding of SU($N$) fermions, our approaches also complement recent ML studies of quantum many-body physics to explore the underlying physics~\cite{Carleo2017,Carrasquilla2017,Deng2017,Zhang:2019hh,Rem:2019hx,Feng:2019ij,Bohrdt:2019fu}.


\vspace{10pt}
\section*{Train neural networks to classify SU($N$) \blue{fermions}}
\vspace{7pt}

We begin by preparing the experimental measurements with appropriate labels. Here, we choose one of the most ordinary experimental measurements for studying SU($N$) Fermi gases, the density profile, and the spin multiplicity as the labels. In our experiment, a degenerate SU($N$) Fermi gas  \blue{with $N$=1,2,5,6} is prepared in an optical trap, and the density profile is recorded by taking spin-insensitive absorption images after time-of-flight expansion, yielding the momentum distribution \blue{The spin multiplicity is confirmed by optical Stern Gerlach measurements (see Methods).} In principle, the density profile contains the momentum-space information of SU($N$) interacting fermions that reflects various thermodynamic observables, \blue{such as Tan's contact or the compressibility,} which is  the underlying reason for the success of using ML techniques to detect the spin multiplicity.   However, the effect of spin multiplicity on the momentum distribution is extremely small compared to other features such as the fugacity and atom number because of small interaction strength. \blue{Therefore, the data set should be prepared in such a way that images are indistinguishable based on the pronounced features (i.e. atom number or temperature), which forces the NN to seek for less pronounced features. We post-select data sets and minimize possible correlations between spin multiplicity and atom number or temperature.} 


In details, we focus on the density profiles with the interaction parameters $k_F a_s\simeq$~0.3 where $k_F$ is the Fermi wave vector and $a_s$ the scattering wavelength and only select the profiles based on similarities in widths of Gaussian fitting of the density profiles to result in indistinguishable momentum profiles as shown in Fig.~\ref{fig:1}b,c (see Methods). We collect 200 density profiles for each class of SU($1$), SU($2$), SU($5$), and SU($6$) (Fig.~\ref{fig:1}b). We randomly feed 150 of them to train the NNs by implementing the supervised machine learning techniques with spin multiplicity as labels, and use the remaining 50 profiles to evaluate the classification accuracy \blue{that is defined as the ratio of number of samples with predictions matching true labels to the total sample number.} To maximize the accuracy of NNs, we choose the architecture of convolutional neural networks (CNNs) that is suited to explore the \blue{less-pronounced effects} in an image (more details in Methods), as shown in Fig.~\ref{fig:1}d.
By choosing the suitable structures and parameters in the CNNs, we can realize a very high accuracy $\sim$94~$\%$, which is much better than the random guess (25~$\%$). We also test various unsupervised learning techniques such as the typical principal component analysis~\cite{PCA1,PCA2} and only get a low classification accuracy of only $\sim$43~$\%$. Moreover, it is worth to emphasize that the remarkably high accuracy $\sim$94~$\%$ of NNs is achieved by using only a single snapshot of the density profiles. All these results indicate that there are \blue{detectable less-pronounced features in a single snapshot of density profile.}

\begin{figure*}[htbp]
	\includegraphics[width=0.85\linewidth]{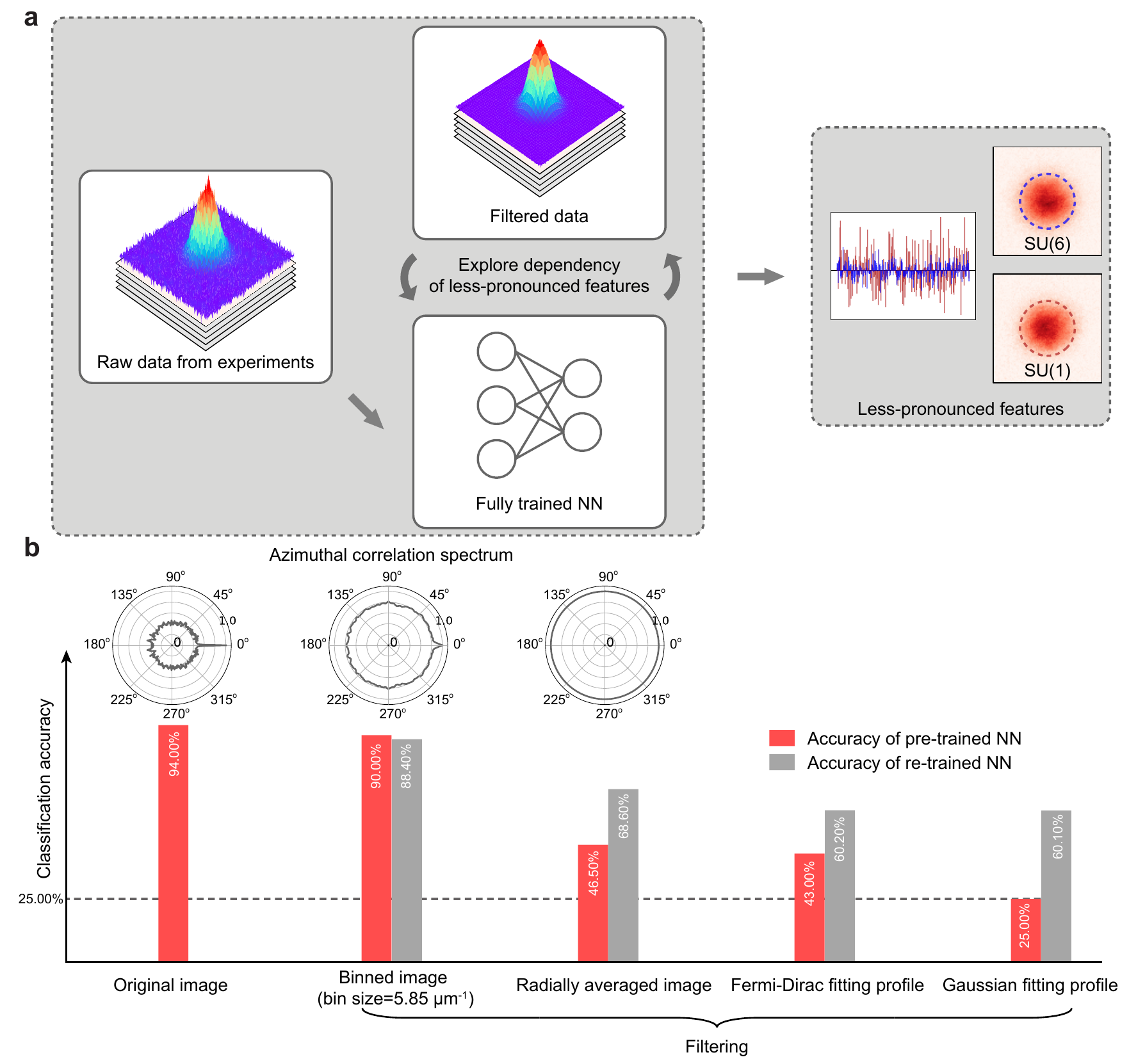}
	\caption{\textbf{\blue{Exploring less-pronounced effects} with NNs using deliberately filtered experimental images} (a) Our heuristic machinery to \blue{uncover less-pronounced effects}. (b) Effect of image filtering on the classification accuracy by neural networks trained with original (red) and filtered (grey) experimental images. The retraining process follows the same procedure of original training with fixed neural network parameters and hyper-parameters (e.g.  the number of epochs and learning rate). In the inset, the azimuthal correlation spectrums calculated at $k=58.5\mu m^{-1}$ are shown.}\label{fig:2}
\end{figure*}

\begin{figure*}[htbp]
	\includegraphics[width=0.85\linewidth]{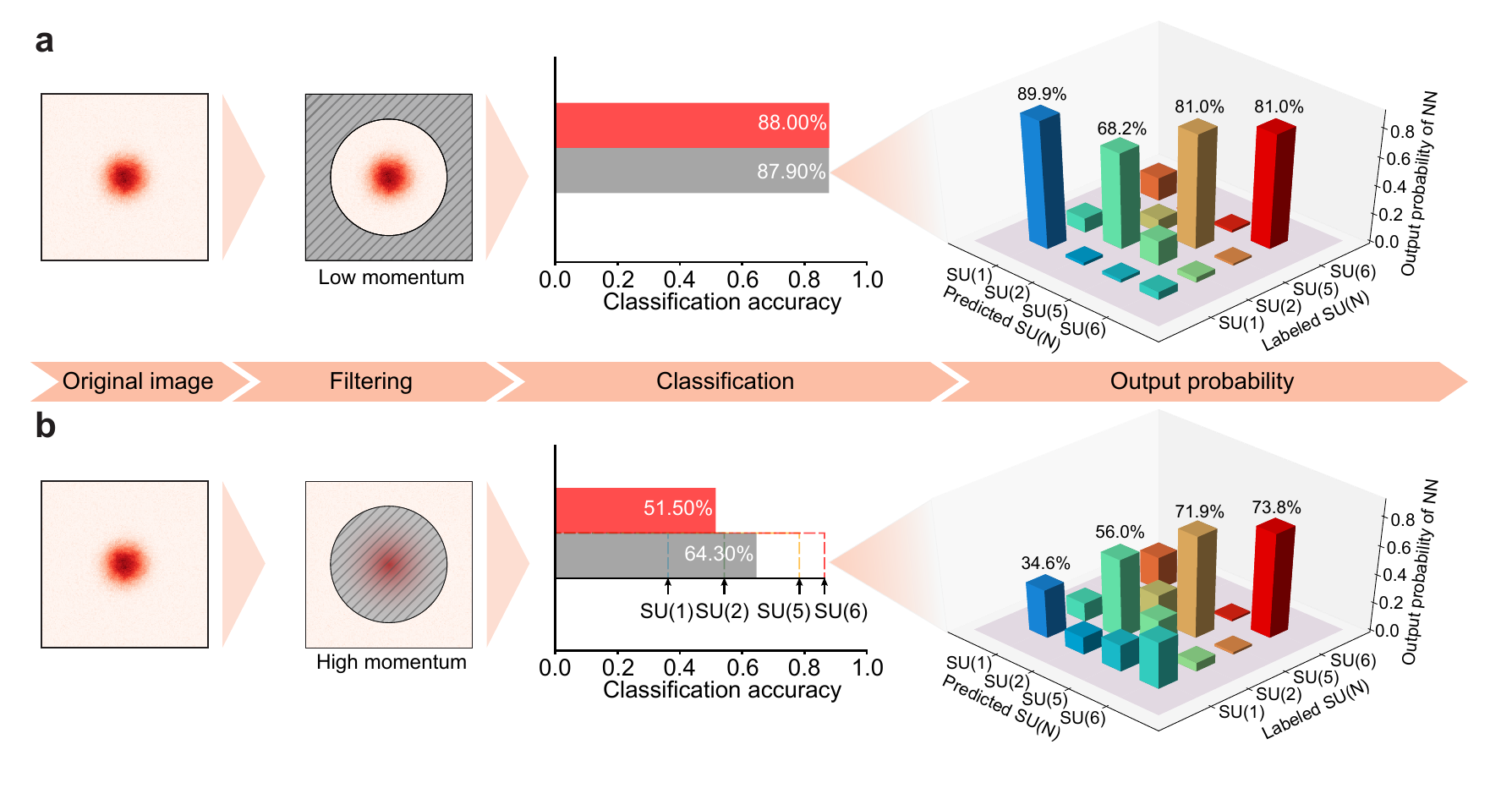}
	\caption{\textbf{\blue{Influence of low and high momentum parts on classification}} Classification accuracy with neural networks trained only with low (a) and high (b) momentum parts. \blue{Red and grey bars represent the accuracy of pre-trained and re-trained NNs, respectively.} Retraining of the neural networks with only high or low momentum parts enhances the accuracy as shown in grey bars.  For high momentum parts, classification accuracy increases with $N$ as shown in black ticks in (b) \blue{where the average of the four investigated SU(N) cases is 64.30\%.} For the re-training NN, the output probabilities of the NN are shown for input images.}\label{fig:3}
\end{figure*}

\vspace{10pt}
\noindent \section*{Extract {\myblue{less-pronounced effects}} in low and high momentum parts}
\vspace{5pt}


We now analyze the attributes processed by the well-trained NNs and extract \blue{less-pronounced effects determined by} the spin multiplicity step by step. Due to the limited \blue{interpretability} of NNs, it is usually difficult to \blue{identify what kinds of features the NNs use for classfication.}
 \blue{In our proposed machinery, we examine which parts of the density profile are related to the spin multiplicity as described in Fig.~\ref{fig:2}a.} Usually, it is more efficient to use some prior knowledge, which can be obtained in our limited understanding of the current system or the well-established understanding of the similar system. In our example of studying the interacting SU($N$) fermions, we  use the prior knowledge of non-interacting fermions and the associated energy (length) scale in choosing various filters in the momentum space. It is conceivable that our heuristic machinery can be applied to other systems.

\blue{To do this, we manually manipulate the experimental images and subsequently  check the classification accuracy of the manipulated images.} 
Since different information is removed in different types of manipulated images, \blue{the classification accuracy will decrease with different amount, which will unfold what kind of information is more important for classification.}
As shown in Fig.~\ref{fig:2}b, we first replace the whole image with the Gaussian and Fermi-Dirac fitting profile to do the test based on the prior knowledge of non-interacting fermions. It turns out that the classification accuracy significantly decreases for both cases, and the accuracy drop of the Gaussian fitting profile is even more, which implies \blue{Fermi-Dirac type preserves the characteristics of the original profile better than the Gaussian fit.}
We further test the variations in accuracies by replacing profiles with radially averaged profiles, which results in test accuracies higher than the Fermi-Dirac fitting cases. However, the differences in accuracies between the radially averaged and Fermi-Dirac are much smaller compared to the differences between Fermi-Dirac and Gaussian, suggesting the SU($N$) dependent modifications of the Fermi-Dirac distribution to be extremely small.


\blue{Now we examine the contributions of low and high momentum parts by classifying the masked images with the well-trained NN as shown in Fig.~\ref{fig:3}a and b. This is motivated by the observation that the classification accuracy significantly decreases with filters with various fitting functions that remove the SU($N$) dependent effect in the high momentum tail.} We choose two different types of masks for the region of the replacements, which will be referred as background and central masks, respectively. Background mask covers from the edge of the image to some atomic momentum $k_c$, while central mask covers from the center to $k_c$. Then we replace the masked region \blue{with a fake image generated by averaging the corresponding region of all the images in the dataset} and re-evaluate the test accuracies of the pre-trained NN.

First, we set the cut-off momentum of $k_c$=70~$\mu m^{-1}$ such that $>$99~$\%$ of atoms are contained within the low momentum region (Fig.~\ref{fig:3}a) that still allows us to classify the spin multiplicity with the accuracy of 88~$\%$. While this confirms that a NN perceive \blue{spin multiplicity dependent} information in low momentum parts, questions remain on why the accuracies are not fully recovered beyond 94~$\%$. Such observation confirms the importance of the high momentum information. In Fig.~\ref{fig:3}b,  we prepare a data set with a high momentum part only ($k>k_c$=70~$\mu m^{-1}$), in which a low momentum region is deliberately replaced by the same fake image. Surprisingly, the test classification accuracy is still $>$50~$\%$, and the overall classification accuracy increases to 65~$\%$ when a NN is re-trained. Such a high accuracy based on the few information of only $<1\%$ of atoms strongly implies that the high momentum tail is crucial for determining the spin multiplicity. \blue{This SU(N)-dependent feature is not due to the finite resolution of the imaging system~\cite{Sanner:2010hq} as the NNs can classify the binned image with high classification accuracy (see Methods for more information).}

\vspace{10pt}
{\bf High momentum tails}
\vspace{5pt}

In light of these results, we speculate that NNs utilize \blue{less-pronounced effects} in the high momentum part. To confirm this, we check the dependance of classification accuracy with the fully trained NNs on each SU($N$=1,2,5,6) class, and we find the classification accuracy increases with $N$ in Fig.~\ref{fig:3}b. In addition, the output probability of the correct spin multiplicity increases with $N$ (see Fig.~\ref{fig:3}b). These results indicate that the \blue{less-pronounced} feature being used in NNs becomes more prominent with increasing spin multiplicity $N$, which is consistent with the fact that atom-atom interactions are absent in the case of SU(1) due to the Pauli principle in the ultracold regime while are significantly enhanced in SU(6) fermions. Indeed, the amount of short-range interactions should be revealed in the high momentum distribution in which the weight of such high momentum tail is determined by the central quantity, so-called the contact~\cite{Tan:2008ey,Tan:2008eg,Tan:2008ch} in a dilute quantum gas. The contact governs many other physical observables~\cite{Stewart:2010fy,Kuhnle:2010je}, and has been probed in strongly interacting gases~\cite{Partridge:2005wb,Kuhnle:2010je,Fletcher:2017cu,Stewart:2010fy,Laurent:2017dp,Yu:2015go,Luciuk:2016gr} and even in a weakly interacting gas with SU($N$) symmetric interactions~\cite{Song:2019wb}. It is conceivable that the NNs detect the high momentum distribution within a single image in contrast to the previous work where hundreds of momentum-space images are statistically averaged in a $^{173}$Yb Fermi gas~\cite{Song:2019wb}. To be noted, our observation is consistent with the direct measurement of the high momentum tail in the region of $k/k_F>3$~\cite{Song:2019wb}, which is corresponding to $k>$100~$\mu m^{-1}$ for an SU(1) gas in this work.

\vspace{7pt}
\textbf{Evaluating detection accuracy with tunable masks}
\vspace{7pt}

\begin{figure*}[htbp]
	\includegraphics[width=0.85\linewidth]{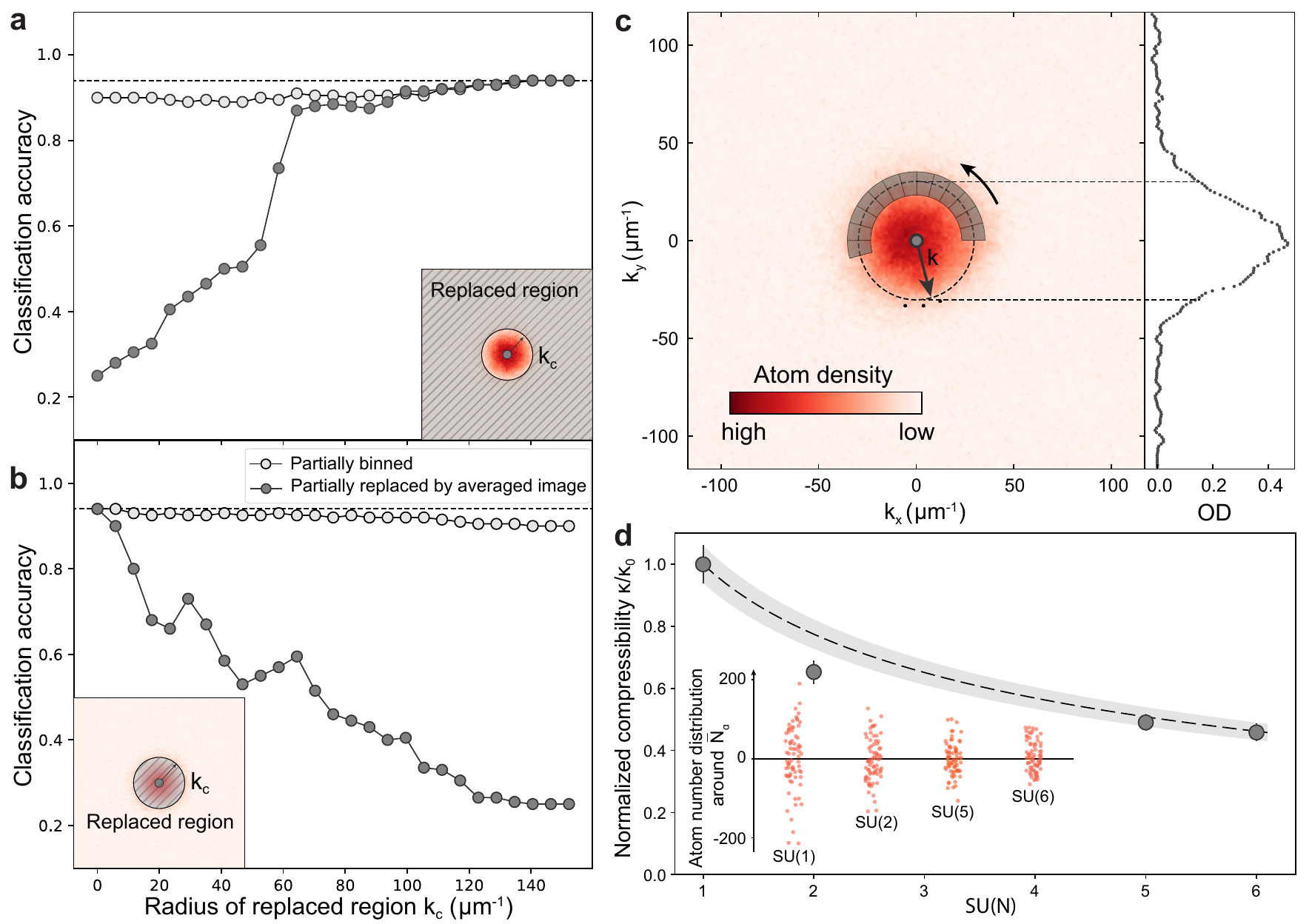}
	\caption{\textbf{Verification of ML-aided detection : density fluctuations and thermodynamic compressibility} (a,b) Classification accuracy of the correct class as a function of the cutoff momentum $k_c$ of the mask. The dotted line indicates the accuracy of 94~$\%$. (c) Measurement of density fluctuations with a snapshot. In a line-of-sight integrated density profile, a series of bins containing on average $\bar{N}_a$ atoms are chosen along the azimuthal direction. Each bin is about 10 (in azimuthal direction) by 17 (in radial direction) $\mu$m$^{-1}$, which is much larger than the optical resolution of the imaging system. The density profile at $k_x$=0 is shown. (d) The normalized compressibility of SU($N$) fermions $\kappa/\kappa_0$ is measured by relative density fluctuations as $\kappa/\kappa_0=\zeta_{SU(N)}/\zeta_{SU(1)}$. The error bar shows the standard error. The dashed line indicates the theory curve $\kappa/\kappa_0=[1+\frac{2}{\pi}k_Fa_s(N-1)(1+\epsilon k_F a_s)]^{-1}$ with the uncertainty represented by the shaded region considering the standard error of $\zeta_{SU(1)}$. \blue{The inset shows the distribution of the atom number per bin from three images for each spin multiplicity. The distribution is plotted around the average normalized by the degenerate temperature,  $(N-\bar{N}_a)/(T/T_F)$, where $\bar{N}_a$ is the average atom number.}}\label{fig:4}
\end{figure*}

To \blue{examine the less-pronounced effects} in both low momentum and high momentum regions and build up the concrete connections between these \blue{less-pronounced effects} and the spin multiplicity, we now quantitatively analyze the changes inflicted on the test accuracies when the cut-off momentum $k_c$ is tuned over. It is clearly shown in  Fig.~\ref{fig:4}a that the test accuracy decreases to $\sim$25$\%$ by complete replacements of the images by the same average image. This accuracy is gradually recovered up to almost 90$\%$ when $k_c$ is increased to $k_c\simeq$70~$\mu m^{-1}$ being consistent with the result in Fig.~\ref{fig:2}, as less information from the low momentum regime is removed. Based on replacement analysis, it is conceivable that SU($N$) dependent features must exist in the low momentum part as we will discuss details below.  In comparison, the classification accuracy of the binned images (bin size=5.85~$\mu m^{-1}$) does not decrease too much, which is only a partial removal of fluctuations. As a complementary study, we utilize the central mask and replace the low momentum information up to variable momenta of $k_c$ in Fig.~\ref{fig:4}b. The classification accuracy gradually decreases with increasing $k_c$ from 0 to $\sim$50~$\mu m^{-1}$ as the information within the density profile is increasingly removed. However, the classification accuracy stays over 50~$\%$ around $k_c=50\sim70$~$\mu m^{-1}$, which strongly suggests that the high momentum tails of the density distribution still contribute towards the classifications based on SU($N$). Beyond $k_c=80$ $\mu m^{-1}$, where the atomic shot-noise becomes comparable to the background shot-noise level, the test accuracy rapidly drops for the images replaced by averaged images.

\vspace{10pt}
\textbf{Density fluctuations and compressibility}
\vspace{5pt}

The question still remains as to what dominant feature classifies spin multiplicity in the low momentum regime. \blue{Based on the significant decrease of the accuracy with profiles being radially averaged in Fig.~\ref{fig:2}b, we hypothesize that the NNs utilize the density fluctuation along the azimuthal direction for classification. The amount of azimuthal density fluctuations can be revealed in the correlation spectrum (Fig.~\ref{fig:2}b) showing a strong signal in the original and binned images while flattened at all angles for the radially averaged images.} 

To understand how the density fluctuations reveal spin multiplicity, we consider the fluctuation-dissipation theorem by which the thermodynamic compressibility $\kappa=\frac{1}{n^2}\frac{\partial n}{
\partial \mu}$ can be measured through density fluctuations (i.e. atom number fluctuations)~\cite{Sanner:2010hq,Mueller:2010ei,Tobias:2020fy,Sonderhouse:2020wz} where $n$ is the local density and $\mu$ the local chemical potential. For repulsively interacting SU($N$) fermions, it is known that the compressibility $\kappa$ decreases with increasing spin multiplicity $N$ as $(\kappa/\kappa_0)^{-1}=1+\frac{2}{\pi}(k_F a_s)(N-1)(1+\epsilon k_Fa_s)$ where $\kappa_0$ is the compressibility of an ideal Fermi gas and $\epsilon=\frac{2}{15\pi}(22-4\ln 2)$~\cite{Yip:2013vr}. Here, the atom number fluctuations are further suppressed by the Pauli blocking in the degenerate regime showing sub-Poissonian fluctuations as $\sigma_{N_a}^2/\bar{N_a}\propto k_B T$ where $N_a$ indicates the atom number measured in the small volume. Therefore, one finds the relative atom number fluctuation $\sigma_{N_a}^2/\bar{N_a}$ is given as
$$\sigma_{N_a}^2/\bar{N_a}=n k_B T \kappa=\frac{3}{2}\frac{T/T_F}{1+\frac{2}{\pi}k_F a_s (N-1)(1+\epsilon k_F a_s)}$$


In our experiment, an atomic sample ballistically expands from the harmonic trap preserving occupation statistics of the phase space during the expansion~\cite{Bruun:2000en,Gupta:2004hy}. Instead of repeatedly producing identical samples and monitoring a small region at the certain position~\cite{Sanner:2010hq}, the relative atom number fluctuations can be extracted along the azimuthal bins containing the same number of atoms on average (therefore resulting in equivalent optical density) within a single image even though a grouping of ideally equivalent bins is challenging and the fluctuation measurement is susceptible to the systematic variations. \blue{The successful classification of the spin multiplicity with NNs now guide us to subsequently investigate the atom number fluctuations with conventional analysis.}

To verify this \blue{less-pronounced} feature, we choose a series of bins containing $\sim$450 atoms on average in a line-of-sight integrated density profile along the azimuthal direction (Fig.~\ref{fig:4}c), and analyze statistics. To have a sufficiently large number of bins for statistical analysis, we perform the measurements at varying momenta on the ring with $(k_x^2+k_y^2)^{1/2}\simeq$28$\mu$m around the center of the distribution. Both the temperature and spin multiplicity are known to affect the atomic density fluctuations through the change in compressibility. Therefore, we normalize relative atom fluctuations by the temperature as $\zeta_{SU(N)}=\frac{\sigma_{N_a}^2}{\bar{N_a}}/\frac{T}{T_F}$ to reveal SU($N$) interaction effect from a single snapshot. In Fig.~\ref{fig:4}d, we then plot the statistical value of $\zeta_{SU(N)}/\zeta_{SU(1)}$. This measurement indeed reveals the \blue{normalized} thermodynamic compressibility $\kappa/\kappa_0=\zeta_{SU(N)}/\zeta_{SU(1)}$ showing the enhanced interaction $k_F a_s (N-1)$. The error bar indicates the standard error from 150 different density profiles.  Whereas the scaling of the measured density fluctuation with $N$ is in good agreement with theoretical prediction, experimental results for SU($N>1$) lie systematically below theoretical ones. The discrepancy may be due to interactions that remain finite during the expansion, which could slightly perturb the occupation statistics of the phase space. \blue{Considering the fact that the change of the compressibility is not significant for $N$=5 and 6 in Fig.~\ref{fig:4}d, the high classification accuracy of NNs using the low momentum part highlights the superior capabilities of single snapshot approach using machine learning. In contrast to the conventional analysis that focuses on a single observable, NNs take a holistic approach in utilizing multiple features simultaneously. Our measurement is in consistent with recent experiments in which thermodynamics is studied by monitoring the density fluctuations and expansion dynamics in degenerate $^{87}$Sr atoms~\cite{Sonderhouse:2020wz}.}

\section*{Discussion and conclusion}

To scrutinize the effects of SU($N$) symmetric interactions, we have provided the NN with altered images and probed specific attributes of the profiles independently. We found that the high momentum tail and density fluctuation information significantly contribute to the SU($N$) classification process. First of all, the high momentum tails of atomic density distributions are expected to exhibit Tan's contact, which encapsulates the many body interactions through the set of universal relations.  While the previous work required averaging of hundreds of images for the detection of the SU($N$) dependent contact~\cite{Song:2019wb}, the NN's ability \blue{makes it possible to obtain the single image distinguishability of the SU($N$) class after training.} However, the exact mechanism behind how the trained network collects the required information for extraction of the contact, whether it is through superior noise suppression or signal enhancement, is not known and left for future work. \blue{Furthermore, it is conceivable that the regression algorithms can be used to extract the change of contact for different spin multiplicity in future works.}

The second dominant feature for the SU($N$) classification is the density fluctuation within the profile. Both the temperature and spin multiplicity are known to affect the atomic density fluctuations through the change in compressibility. Sub-Poissonian density distributions have been observed in degenerate Fermi gases of atoms~\cite{Sanner:2010hq,Mueller:2010ei} and molecules~\cite{Tobias:2020fy}, where multiple images were used to obtain the statistics. \blue{The suppression of the density fluctuation was also observed in SU($N$) fermions allowing for the thermodynamic study~\cite{Sonderhouse:2020wz}.} For a single image, there exists multiple sets of density fluctuation measurements at varying \blue{momentum}, where each measurements form a ring around the center of the distribution. Considering the decreased SU($N$) classification accuracy from the radially averaged data sets, the fluctuation information might have been utilized in addition to the contact to reflect the effects of compressibility. Lastly, we found that the low energy part of the density profile does not exhibit a signature as strong as the previous two features. While there has been a report of SU($N$) dependent modifications to the density distribution limited to the 1D case~\cite{Pagano:2014hy}, the corresponding beyond mean-field effects in 3D remains challenging to be measured experimentally.


In conclusion, we have demonstrated the capabilities of the proposed machinery by classifying SU($N$) Fermi gases with their time-of-flight density distributions. The neural network provides classifications with an accuracy well beyond the conventional methods such as PCA. By applying different types of manipulations, we also find that the NNs combine the features from a high momentum signal and density fluctuations together to distinguish SU($N$). Future directions include predictions of $T/T_F$ of SU($N$) Fermi gases based on regression algorithms, and explorations of human feedbacks to the machine learning process for feature extractions. \blue{Feature extraction through machine learning may guide us to investigate the right information and facilitate research in many-body quantum systems.} 
\vspace{10pt}


\vspace{.1in} \noindent
\textbf{DATA AVAILABILITY}
The data that support the finding of this work are available from the corresponding authors upon request.
\vspace{.1in} 

\noindent \textbf{CODE AVAILABILITY}
The code developed during the current study is available from the corresponding authors upon request.
\vspace{.1in} \\

\noindent\textbf{ACKNOWLEDGMENTS}
We thank Tin-Lun Ho for helpful discussion. G.-B.J. acknowledges supports from the RGC and the Croucher Foundation through 16311516, 16305317, 16304918, 16306119, C6005-17G and N-HKUST601/17. This work is partially supported by the Innovative Exploratory Grant at HKUST. J.L. acknowledges supports from the RGC (N-HKUST626/18, 26302118 and 16305019).
\vspace{.05in} \\

\vspace{.1in} \noindent
\textbf{COMPETING INTERESTS}
The authors declare that they have no competing interests.
\vspace{.05in} \\

\vspace{.1in} \noindent
Corresponding author : gbjo@ust.hk\\

\clearpage
\vspace{.1in} \noindent
\textbf{METHODS}
\vspace{.05in} \\

\noindent \textbf{Sample preparation} We prepare a balanced ultracold Fermi gas of $^{173}$Yb atoms with SU($N$) symmetric interactions as large as $N$=6. (shown in Fig.\ref{fig:1}a). We begin by loading a laser cooled six-component Fermi gas, where the nuclear spin states are equally populated, into a three-dimensional optical dipole trap (ODT). The atoms are further evaporatively cooled in the ODT to a temperature range of 0.2 to 1.0 $T/T_F$, where $T_F$ is the Fermi temperature. During the evaporation, different spin configurations are prepared via an optical pumping process using a narrow line-width transition of $^1S_0(F=5/2)\rightarrow$ $^3P_1(F'=7/2)$ at a wavelength of $\lambda=556$~\blue{nm}. The $\sigma^\pm$-polarized pumping light removes unwanted $m_F$ states of the ground manifold of $^1S_0$~\cite{Song:2016ep} and produces a Fermi gas with tunable SU($N$) interactions, as the nuclear spin relaxation rates are negligible in our experiment. After the evaporative cooling, the ODT is further ramped up in 60~ms to obtain large enough trap frequencies $(\omega_x,\omega_y,\omega_z)$$=2\pi\times$$(1400,750,250)$~Hz before 4 ms of time-of-flight expansion. We measure the density distributions by taking absorption images using a spin-insensitive $^1S_0(F=5/2)\rightarrow$ $^1P_1(F'=7/2)$ transition at 399 nm. The images are taken in random order with respect to their spin configurations, to avoid the possibility of a classification based on fluctuations in the background. \blue{The spin configuration of the sample can be monitored by the optical Stern Gerlach measurement. In general, the atom number of different spin states has a  fluctuation of $\pm 2\%$ of the total atom number. }


\vspace{10pt}

\noindent \textbf{Data preparation}  All snapshots are first preprocessed by the fringe removal algorithm reported in Ref.~\cite{Song:2020wn}. Then cropped images are loaded into the neural network for further classification. For SU($N$) data, it is natural to prepare the same number of atoms per spin at constant $T/T_F$, in which the normalized density profile is the same for different SU($N$) cases. In this case, however, we find that the diffraction of the imaging light induces fringe patterns that depend on the total atom number in the experiment. \blue{One can normalize the image by the total atom number, but we inevitably change the level of background noise. Therefore, we keep the total atom number unchanged otherwise the neural network uses the background fringe patterns or noises to classify the SU($N$) data. In our experiment, we post-select 200 images per each SU($N$) class by using a Gaussian fitting, which allows us to obtain samples with similar profiles but different $T/T_F$. If we have kept different SU($N$) gases at constant $T/T_F$, the profiles are identical in the unit of $k_F$, instead of  pixel. Subsequently, 75 percent of the data is used for training NNs and the remaining is for test.}

\vspace{10pt}

\noindent \blue{\textbf{Machine learning} Machine learning, a sub-field of artificial intelligence, allows us to understand the structure of data and deduce models that explain the data. Traditionally, machine learning can be classified into two main categories: supervised and unsupervised learning, based on whether there are labels or not for training. Supervised learning usually trains a model from a known dataset of input data $\{x_i\}$ and output labels $\{y_i\}$ so that the model can find a correspondence rule $x_i\mapsto y_i$ which allows us to predict the labels of data beyond the training dataset. In contrast, unsupervised
learning is used to classify the data into several different clusters based on the potential patterns or intrinsic structures in the dataset without any priori knowledge of the system or data properties. \\}

\noindent \blue{\textbf{Convolutional neural network} Machine learning techniques used in this work are based on  convolutional neural networks (CNNs), which takes a supervised learning approach for classification task. Neural networks (NN), inspired by the biological neural networks that constitute animal brains, are composed of a series of artificial neurons, among which the connection is a real-valued function $f: \mathbb{R}^k \rightarrow \mathbb{R}$, parametrized by a vector of weights $(w_1, w_2, ..., w_i, ...)=\bm{w}\in\mathbb{R}^k$ and the activation function $\phi: \mathbb{R}\rightarrow\mathbb{R}$, given by}

\blue{$$ f(x)=\phi(\bm{w}\cdot \bm{x})\quad{\rm with}\quad\bm{x}=(x_1,...,x_i,...)\in \mathbb{R}^k.$$}

\noindent \blue{By combining the artificial neurons in a network or in a layer of network, we obtain NNs. In recent years, CNNs have shown stronger validity and better performance than regular NN in image recognition. Similar to the regular NNs, CNNs also consist of a sequence of layers and each layer receives some inputs, performs a dot product and optionally follows it with a non-linear activation function.  However, unlike a regular neural network, a CNN usually has several convolutional layers where neurons are arranged in 2 dimensions providing an efficient way of detecting the spatial structure. The convolutional layer first accepts an input 2D image from the previous layer or the whole neural network. {Then the kernel of the convolutional layer slides (i.e. convolve) across the width and  height of the input image with dot products between the kernel and the input being computed. Consequently, we obtain a 2D feature map in which each pixel is the response at corresponding position. If the convolutional layer has $N$ different kernels, the same procedure will be repeated for each kernel and finally $N$ 2D feature maps will be produced.} These 2D feature maps will then be loaded into the next layer as input. \\}

\noindent \blue{\textbf{Training and evaluating the NN} The CNNs used in this paper are realized by using the Tensorflow in Python~\cite{tensorflow}. We have attempted different architectures and found that the result is not sensitive to the choice of architecture such as the number of layers or the kernel size. Therefore, we remove superfluous layers to simplify our model. The concrete parameters taken in this work  are listed in Table \ref{tab:1}. }

\begin{table}[htbp]
	
	\resizebox{0.48\textwidth}{24mm}{
		\begin{tabular}{l|l|l|l}
			Layer & Layer name        & Function              & Description                                                                               \\ \hline
			1     & Input             & Image input           & 201$\times$201 images                                                                     \\
			2     & Conv. layer       & Convolution           & \begin{tabular}[c]{@{}l@{}}24 24$\times$24 convolutions\\  with stride (1,1)\end{tabular} \\
			3     & ReLU              & Activation function   & ReLU function                                                                             \\
			4     & Pool layer        & Average pooling       & 2$\times$2 with stride (1,1)                                                              \\
			5     & Dropout           & Dropout               & 50\% dropout                                                                              \\
			6     & Fully conn. layer & Fully connected       & \begin{tabular}[c]{@{}l@{}}Fully connected layer\\  with 4 neurons\end{tabular}           \\
			7     & Softmax           & Activation function        & Softmax function                                                                          \\
			8     & Output            & Classification output & \begin{tabular}[c]{@{}l@{}}Probability with classes\\  "N=1,2,5,6"\end{tabular}          
	\end{tabular}}
	\caption{Network architecture and parameters used in this work}	\label{tab:1}
\end{table}

\blue{To train the network on the data with different spin configurations, the model is compiled with a cross-entropy loss function. During the training process, the weights of model are updated based on Adam algorithm \cite{Kingma2014adam} to minimize the loss function with a learning rate of $1\cdot 10^{-4}$, which is a hyper-parameter that controls how much the network changes the model each time. The maximum training epochs are limited to 1000 and the accuracy and loss are monitored during the training process for selecting the model with best performance. After full training, we evaluate the trained model on the test dataset.}

\blue{We characterize the performance of trained NNs by obtaining the overall classification accuracy, which is defined as the ratio of number of samples with predictions matching true labels to the total sample number. For one single image loaded into the NNs, for example, the softmax activation function normalizes the output values $\{\sigma_c\}$ by $P(c)=e^{\sigma_c}/\sum_{c=1}^4e^{\sigma_c}$, which allows a probabilistic interpretation for the different classes denoted by the subscript c. When calculating the classification accuracy, the class with highest probability $P(c)$ is selected as the prediction from NNs. As a complementary analysis of NNs, we also evaluate an output probability matrix that hints how the NNs perform among different classes, like shown in Fig.1d and Fig.3. In the probability matrix, every element $A_{ij}$ represents the probability $P(c)$ averaged over the results for all images with the true label $j$ and prediction $i$.\\}

\vspace{10pt}

\begin{figure}[htbp]
	\centering 
	\includegraphics{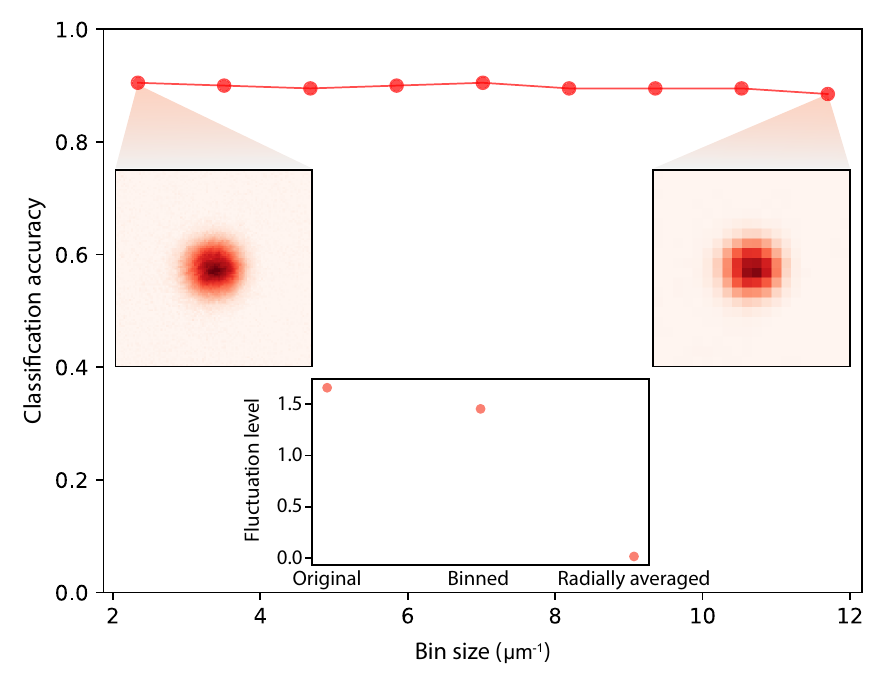}
	\caption{\textbf{Classification accuracy with different bin size.} The classification accuracy is quite robust when bin size $n$ is changed from $2.34\mu m^{-1}$ to $11.70\mu m^{-1}$. The inset shows the comparison of fluctuation level among original image, binned image and radially average image.}\label{fig:5}
\end{figure}

\noindent \blue{\textbf{Manipulation of SU($N$) data} In this work, we manipulate the experimental images to remove different types of information. In Fig.~\ref{fig:2}, we examine the binned image, radially averaged image, Fermi-Dirac fitting profile and Gaussian fitting profile. The blurring of adjacent pixels effectively changes subtle features in SU($N$) gases due to finite optical resolution , for example it will decrease the measured atom number variance~\cite{Sanner:2010hq}. We minimize this effect by binning the data using a sufficiently large bin size~\cite{Sanner:2010hq}.}  {\color{blue}The whole image is partitioned into bins with the area of $n\ \mu m^{-1}\times n\ \mu m^{-1}$ without overlapping. In each bin, the averaged optical density with the bin is calculated, and the value is subsequently used to fill all the pixels of the bin to maintain the original size of image.} We attempted several different bin size $n$ from $2.34\mu m^{-1}$ to $11.70\mu m^{-1}$ and the result is robust against the bin size, as shown in Fig.~\ref{fig:5}. 

\blue{For the radially averaged images, we first divided all the pixels into several bins based on the distance from the center of the atom cloud, then  averaged the pixels in the same bin. The degenerate Fermi-Dirac and Gaussian profiles are fitted by the 2D Thomas-Fermi and Gaussian distribution, respectively. It is worth to note that both density fluctuations and high momentum information are effectively removed from both fitting cases. Therefore, the comparison between Fermi-Dirac and Gaussian profiles may allow one to investigate possible next order effects by which NNs detect the changes in $T/T_F$.}

\blue{In Fig.~\ref{fig:3}, we first divide the whole image into two parts, low momentum and high momentum regions, based on whether the distance from the center of the atom cloud is larger than 70 $\mu m^{-1}$ or not. Next, we replace one of the two parts with a fake image, which is generated by averaging the corresponding region of all the images ($N$=1,2,5,6) in the dataset. Since all the test images are  same in the replaced region, the information in that region can be considered as removed. In Fig.~\ref{fig:4}a and b, the procedure is the same as Fig.~\ref{fig:3} with variable  cut-off momenta. For the partially binned images, the corresponding region is replaced by a binned image.}

\vspace{10pt}
\noindent \textbf{Data analysis} In Fig.~\ref{fig:2}a, we calculate the azimuthal correlation spectrum at  $k=58.5\mu m^{-1}$, which is defined as $C_k (\theta_j)=\frac{\sum_i[P_k(\theta_i)P_k(\theta_{i+j})]}{\sum_iP_k^2(\theta_i)}$, where $P_k(\theta_i)$ represents the optical density for a specific pixel at $k\sim 58.5\mu m^{-1}$ and angle $\sim \theta_i$. The formula can be further derived from the Fourier transform $\mathcal{F}$ as $G=\frac{\mathcal{F}^{-1}(|\mathcal{F}(P_k)|^2)}{|P_k|^2}$. The azimuthal correlation spectrum shows how the image looks like its own copy after rotating a specific angle. For a radially averaged image, the correlation becomes 1 at any angle indicating no density fluctuations. When the image is binned, densities are only locally averaged resulting azimuthal correlation at nonzero angle less than 1.

Fig.~\ref{fig:4}d shows density fluctuations measured along azimuthal bins containing the same number of atoms within a single image. Total 24 bins are chosen at the distance of $22\sim34\mu m^{-1}$ from the center of the cloud. Redundant pixels are removed at the border to keep all bins have the same number of pixels. The size of each bin is much larger than the optical resolution of the imaging system.

\end{document}